\documentclass{elsart}
\usepackage{epsfig}

\begin{document}

\begin{frontmatter}

\title{The intercept of the BFKL pomeron from Forward Jets at HERA}

\author[Merida]{J. G. Contreras\thanksref{email}} 

\address[Merida]{Departamento de F\'{\i}sica Aplicada,
CINVESTAV--IPN, Unidad M\'erida, A. P. 73 Cordemex, 97310 M\'erida, 
Yucat\'an, M\'exico}
 
\thanks[email]{email: jgcn@monika.cieamer.conacyt.mx}

\begin{abstract}
  Recently the H1 and ZEUS collaborations have presented 
  cross sections for DIS events with a forward jet. 
  The BFKL formalism is able to produce an excellent fit to these
  data. The extracted intercept of the hard pomeron suggests that when
  all higher order corrections are taken into account the cross
  section will still rise very rapidly as expected for low $x$ dynamics.
\end{abstract}

\end{frontmatter}

PACS: 13.60, 13.87

Keywords: BFKL, DIS, Forward Jets

\section{Introduction}

\begin{figure}
\centering
\epsfig{file=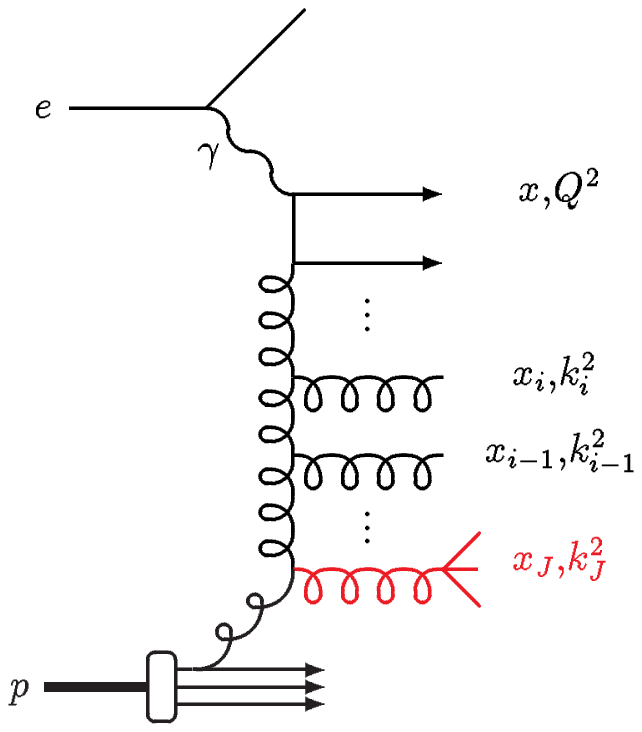, height=.7\textheight}
\caption{Representation of a deep inelastic event using a physical
  guage within perturbative
  QCD.}
\label{f:diagram}
\end{figure}

Using a physical gauge, deep inelastic scattering processes can be
represented within perturbative QCD with ladder diagrams like the one
shown in figure \ref{f:diagram}. In this picture the interaction
between an electron and a proton is mediated through the exchange of a
virtual photon of four momentum squared $q_\mu q^\mu=-Q^2$, which
couples to a quark--antiquark box at the top of the parton ladder
inside the proton.

The DGLAP \cite{dglap} and BFKL \cite{bfkl} schemes select different
leading logarithmic regions of the phase space to describe the
partonic evolution along this ladder. The DGLAP approach takes into
account the leading terms in $\ln Q^2$, but neglects those in $\ln
(1/x)$. In this case the square of the transverse momentum of the
partons along the ladder are strongly order, $k^2_i\gg k^2_{i-1}$
({\em cf.} fig. \ref{f:diagram}), whereas the longitudinal momenta
obey $x_i<x_{i-1}$.  On the other hand, in the BFKL formalism the $\ln
(1/x)$ are resummed in the region of $k^2_i\approx k^2_{i-1}$ and
$x_i\ll x_{i-1}$. Due to the inherent approximations involved the
domain of applicability of the DGLAP equations is medium to high
$x$--Bjorken, but the BFKL approach should only be used at small
$x$--Bjorken.
 
The electron--proton collider HERA opened up the study of 
DIS to new domains of small values of $x$--Bjorken
and still sizeable virtuality of the photon. It was thus
expected that experiments at 
HERA would be able to measure the transition from the
region of applicability of DGLAP to that of BFKL. 
Among the different proposed observables to test the BFKL approach to
DIS, the measurement of forward jets is considered to be one of the most
promising.  

Higher order corrections to the BFKL formalism have been
recently calculated~\cite{lipatov}.
They turned out to be sizable and pointed out 
the need of still higher order corrections. This reduced the
quantitative predictive power of the leading logarithmic approximation to
 BFKL calculations, but the
qualitative behavior at small $x$ may still remain the same. Taking
this into account a fit to HERA data, which of course resumes all
orders, may shed light into the problem of the numerical stability of
the BFKL kernel under higher order corrections. In this spirit the
intercept of the BFKL pomeron is extracted from a fit to the
forward jet data of the H1 and ZEUS collaborations.

\section{Forward jets at HERA}

Back in 1990 A. H. Mueller \cite{mueller} proposed to look for DIS
events with one jet --other than the jet originating from the struck
quark-- fulfilling the following characteristics ({\em cf.} fig.
\ref{f:diagram}):

\begin{itemize}
\item  $\mathbf{x}$ {\bf small}. The selection of the smallest
  $x$--Bjorken experimentally possible implies going away from the
  domain of validity of DGLAP and into a phase space area governed by BFKL.
\item $\mathbf{x_J}$ {\bf large}. This, along with the previous item,
  provides the phase space for parton evolution and has the extra
  advantage to enable the use of parton density functions (PDF) of the
  proton in a region where they have already been measured, so no
  extrapolation is needed.
\item $\mathbf{k^2_J\approx Q^2}$. This requirement suppresses 
  DGLAP evolution without affecting the
  BFKL dynamics. It also provides a big scale at both ends of the
  ladder, avoiding thus the dangerous infrared regions of phase space
  and increasing the solidity of the analytic predictions.
\end{itemize}

The process so selected provides a well defined topology for the
experimentalist. One has to look for a DIS event with a jet at high
rapidities and enforce that the virtuality of the proton is of the
same scale as the transverse momentum squared of the jet. As the
direction of the proton is termed by the HERA experiments the forward
direction, this kind of events are known as forward jets events.
Recently both the H1 \cite{h1} and the ZEUS \cite{zeus} collaborations
have presented cross sections for this observable. The H1
collaboration selects DIS events requiring the energy ($E_e$) and polar
angle ($\theta_e$) of the scattered electron to be $E_e>11$ GeV and
$160^\circ<\theta_e<173^\circ$. ZEUS only requires $E_e>10$ GeV and
the polar angle $\theta_e$ is restricted by the rest of the kinematic
constrains. Both of them select events with $y_{Bj}>0.1$ to avoid the
jet from the struck quark going forward. In the jet side, selected
with a cone algorithm of radius 1, the cuts used are for H1: $k_J>3.5$
GeV, $x_J\approx E_J/E_{beam}=E_J/820$GeV$>0.035$ and
$7^\circ<\theta_J<20^\circ$. ZEUS ask for jets with $k_J>5.0$ GeV,
$x_J>0.036$ and $8.5^\circ<\theta_J$. Both collaborations require that
$0.5<k^2_J/Q^2<2.0$. The luminosity used is 2.8 and 6.4 pb$^{-1}$ for
the H1 and ZEUS collaborations respectively.  The H1 collaboration
also reports a similar search, where all cuts were kept the same,
except that $k_J>5.0$ GeV was required.  The measured cross sections
for different bins in $x$--Bjorken are presented in the first column
of tables \ref{t:cs} and \ref{t:cs1}. Note that both collaborations
presented slightly asymmetrical systematic errors. The errors presented
in tables \ref{t:cs} and \ref{t:cs1} are an average of the systematic
errors added in quadrature to the statistical error of the
measurement.  Note also that the measurements at the lowest
$x$--Bjorken have not been included. This is because, due to the
$0.5<k^2_J/Q^2<2.0$ cut, the experiments ran out of phase space. Thus
these points mix dynamic effects with phase space restrictions and are
not useful for the analysis presented here.

\section{BFKL fits to the forward jet data} 
 
The cross section for DIS events containing a forward jet has been
calculated at leading logarithmic approximation
 within the BFKL formalism in references
\cite{bdm,tang,kms}. There the following form has been found:

\begin{equation}
k^2_Jx_J\frac{d^4\sigma}{dxdQ^2dx_Jdk^2_J} = C \alpha_s(Q^2)
F(x_J,Q^2)\sqrt{\frac{Q^2}{k^2_J}} 
\frac{\exp[(\alpha_p-1)\ln x_J/x]}{(\ln x_J/x)^{1/2}}
\label{e:bfkl}
\end{equation}

where $\alpha_s(Q^2)$ is the QCD coupling constant at the scale $Q^2$
and $F$ is a generic parton distribution in the proton given by

\begin{equation}
F(x_J,Q^2)=x_JG(x_J,Q^2)+\frac{4}{9}[x_Jq(x_J,Q^2)+x_J\bar{q}(x_J,Q^2)].
\end{equation}

In \cite{bdm,tang,kms} an explicit form for the parameters $C$ and
$\alpha_p$ can be found. Recently the NLO corrections to the BFKL
kernel have been presented \cite{lipatov}. These corrections turn out
to be very large implying that NNLO calculations are needed. In
particular the corrections affected the intercept of the BFKL pomeron
$\alpha_p$, reducing
the predictive power of the explicit forms given in
\cite{bdm,tang,kms} at LO and in \cite{lipatov} at NLO.
This means that the exact power of the leading behavior of equation
(\ref{e:bfkl}),

\begin{equation}
(\frac{x_J}{x})^{\alpha_p-1},
\label{e:power}
\end{equation}
 
is not numerically known from BFKL calculations. Nonetheless given
that the first experimental data for this process are available, it is
tempting to test if the form of equation (\ref{e:bfkl}) does indeed
describe the measurement, and if so, which intercept of the BFKL
pomeron is favored by the data.

To perform the fit there are some other ingredients needed. Standard
parton density functions are required to evaluate $F$. Also as
equation (\ref{e:bfkl}) is a four differential expression, values for
$x$, $Q^2$, $x_J$ and $k^2_J$ are needed in each $x$--Bjorken bin.
Both collaborations, H1 and ZEUS, report that the Monte Carlo
generator Ariadne \cite{ariadne} describes very well, not only the
$x$--Bjorken dependence, but all distributions involved in the
analysis. So a similar analysis to that reported by both
collaborations has been performed at the hadron level of the Ariadne
Monte Carlo. It has been checked that the cross section obtained with
this procedure agrees with both, the reported data and also with the
expectations from the Ariadne Monte Carlo as given by the H1 and ZEUS
collaborations. Using this Monte Carlo data set for forward jet events
the mean values of the variables $x$, $Q^2$, $x_J$ and $k_J$ have
been estimated for each measured point of both collaborations. The
results are presented in tables \ref{t:cs} and \ref{t:cs1} along with the
measured data points and their errors.
  
Using the input of tables \ref{t:cs} and \ref{t:cs1}, the PDFs
given by GRV-LO \cite{grvlo}, GRV-HO \cite{grvho}, CTEQ-4M \cite{cteq}
and MRS-R1 \cite{mrsr1} --last three calculated in the
$\overline{\mbox{MS}}$ scheme--, and an $\alpha_s$ value consistent with
each of the different parton density functions, a 2 parameter
fit was performed separately to the H1 and the ZEUS points using formula
(\ref{e:bfkl}).  The $\chi^2$ and the value of $\alpha_p$ obtained are
shown in table \ref{t:fit}.

\section{Discussion}

The results of the fits can be summarized as follows:

{\bf 1.} All data points could be successfully fit to the form of
 equation (\ref{e:bfkl}). This is quite encouraging because that was
 the main motivation for this kind of measurements.
 
 {\bf 2.} The fits are insensitive to the PDF used. This result is
 also as expected. As a matter of fact this was one of the main
 advantages of the forward jet proposal.

 {\bf 3.} Different values of the exponent $\alpha_p$ are found when
 using LO or NLO PDF. On the one hand, being the formula (\ref{e:bfkl}) a
 LO approximation to BFKL, one is tempted to consider only the use of
 LO PDFs. On the other hand, the whole idea of performing the fits is
 to have a data driven estimation of the effects of higher order
 corrections to the BFKL kernel. The actual variation of the value of
 $\alpha_p$ is expected in the basis of it being proportional to
 $\alpha_s$ in LO. As the value of $\alpha_s$ decreases in going from
 LO to NLO, the value of $\alpha_p$ must compensate this trend and
 increase from one case to the other.

 {\bf 4.} The values of $\alpha_p$ found using H1 data are different to
 those found using ZEUS measurements. On the one hand ZEUS data is more
 precise having a factor of 3 more luminosity. This allowed the ZEUS
 collaboration to measure the cross section in a region definitely
 dominated by DGLAP, having thus a clean transition from the box
 diagram at the top of figure \ref{f:diagram} to the case when the box
 is complemented with a ladder. This greatly constrains the fit to equation
 (\ref{e:bfkl}).  On the other hand H1
 points reach smaller $x$, which is the interesting region for BFKL
 studies, although still with huge errors. One possible explanation for
 the difference in the values obtained using the data of both
 collaborations,  could be again the
 $\alpha_s$ dependence of $\alpha_p$. Note that the average $Q^2$ is
 a lot bigger in the ZEUS measurement than in the H1 case. In the LO
 approximation this would naively produce a 15 to 20\% increase of
 $\alpha_p$ from H1 data with respect to that from ZEUS cross sections. 
 
 {\bf 5.} The formula (\ref{e:bfkl}) is at parton level, i.e. it does
 not includes hadronization effects. As reported by both collaborations
 these are uncertain. Different models yield not only different
 normalization, but may also yield  a $x$ dependence (see for example
 figure 8 on reference \cite{zeus}) of the correction from hadron to
 parton level.

 {\bf 6.} Using the dipole approach to BFKL, it has been shown that
 the measurements of the structure function $F_2$ can be described
 using an intercept of $\alpha_p=1.28$ \cite{royon}. Some care is
 necessary when comparing this value with the one obtained here. The
 question of infrared divergencies due to the random walk generated
 by the BFKL kernel \cite{lotter} is quite sensitive for the $F_2$
 case, but it does not appear in the case of forward
 jets. Nevertheless is quite comforting that two so different
 observables yield results compatible with the BFKL formalism.

 {\bf 7.} The possibility to experimentally reach lower $x$ at still
 sizable $Q^2$ is very important. The fits presented here, although
 they could not used the lowest $x$ points due to phase space
 constrains, have shown that
 the BFKL dynamics enforces a steep rise of the cross section for a
 process governed by a hard pomeron inspite the NLO corrections to the
 BFKL kernel. Reaching smaller $x$ will allow
 to  access the saturation region of hot spots in the proton, which
 was one of the primary motivations of the forward jet proposal. This
 goal may still be reachable at HERA.

\section{Conclusion}

A fit was performed of a BFKL prediction to forward jet production as
measured by the  H1 and ZEUS collaborations. All data were consistent with
the assumption of using the BFKL formula for this process. Difference
in the intercept of the pomeron obtained with different sets of data,
may be assigned to its $\alpha_s$ dependence. The fits support the
idea of a hard pomeron and point in the direction that when all higher
order corrections are taken into account the forward jet
cross section will still be rising quite rapidly.

\listoffigures
 
\begin{table}
\centering
\caption{Cross sections for the forward jet production measured by
the H1 \cite{h1}  collaboration along with
average values for the kinematic variables obtained from the Ariadne
MC \cite{ariadne}. Note that the quoted errors are an average of the
slightly asymmetric cuts reported by the collaboration.}
\begin{tabular}{ccccc}
\hline
\hline
$\sigma$ [nb] & $<x>$ & $<Q^2>$ [GeV$^2$] & $<E_J>$ [GeV] & $<k_J>$ [GeV]\\
\hline
\hline
\multicolumn{5}{c}{H1 $k_J>3.5$ GeV} \\
\hline
342 $\pm$ 55 & 0.00073 & 21.5 &34.4 &5.0 \\
224 $\pm$ 32 & 0.0012 & 26.9 &35.4 & 5.5 \\
138 $\pm$ 25 & 0.0017 & 31.4 & 36.9&5.8 \\
67 $\pm$ 11 & 0.0024& 38.1&38.1 &6.3 \\
32 $\pm$ 5 & 0.0035 &47.0 &38.8 &6.9 \\
\hline
\multicolumn{5}{c}{H1 $k_J>5.0$ GeV} \\
\hline
132 $\pm$ 20 & 0.0012 &32.2 &37.9 &6.3 \\
96 $\pm$  20 & 0.0017 &334.8 &39.3 &6.5 \\
55 $\pm$ 10 & 0.0024 &40.1 &39.4 &6.7 \\
28 $\pm$ 6 & 0.0035 & 48.2&39.6 &7.2 \\
\hline
\hline
\end{tabular}
\label{t:cs}
\end{table}

\begin{table}
\centering
\caption{Cross sections for the forward jet production measured by
the ZEUS \cite{zeus} collaboration along with
average values for the kinematic variables obtained from the Ariadne
MC \cite{ariadne}. Note that the quoted errors are an average of the
slightly asymmetric cuts reported by the collaboration.}
\begin{tabular}{ccccc}
\hline
\hline
$\sigma$ [nb] & $<x>$ & $<Q^2>$ [GeV$^2$] & $<E_J>$ [GeV] & $<k_J>$ [GeV]\\
\hline
\hline
77.8 $\pm$ 7.6 & 0.0019 &50.7 &39.9 &7.6 \\
34.4 $\pm$ 3.6 & 0.0033 & 75.6&43.8 &8.7 \\
14.1 $\pm$ 2.1 & 0.006 & 113.6&49.6 &10.4 \\
6.5 $\pm$ 0.7 & 0.010 & 176.4& 58.5&12.9 \\
2.7 $\pm$ 0.4 & 0.018 & 244.7&67.3 &15.1 \\
0.6 $\pm$ 0.3 & 0.031 & 366.8&78.8 &18.8 \\
\hline
\hline
\end{tabular}
\label{t:cs1}
\end{table}

\begin{table}
\centering
\caption{Values obtained for $\alpha_p$ from a fit of the BFKL
  formalism to the data on forward jet production using different PDFs.}
\begin{tabular}{ccccccc}
\hline
\hline
PDF & \multicolumn{2}{c}{H1 $k_J>3.5$ GeV} &\multicolumn{2}{c}{H1
  $k_J>5.0$ GeV} &  \multicolumn{2}{c}{ZEUS}\\
\hline
\hline
& $\chi^2$ & $\alpha_p$ & $\chi^2$ & $\alpha_p$ & $\chi^2$ & $\alpha_p$ \\
\hline
GRV--LO & 1.05 & 1.6 $\pm$ 0.4 & 0.24 & 1.7 $\pm$ 0.5 & 0.81 &
1.15 $\pm$ 0.13 \\
GRV--HO & 1.0 & 1.8 $\pm$ 0.4 & 0.24 & 1.8 $\pm$ 0.5 & 0.76 & 1.24
$\pm$ 0.14\\
CTEQ--4M & 0.99 & 1.8 $\pm$ 0.4 & 0.22 & 1.8 $\pm$ 0.5 & 0.79 & 1.26
$\pm$ 0.14\\
MRS--R1& 1.0 & 1.8 $\pm$ 0.4 & 0.24 & 1.8 $\pm$ 0.5 & 0.91 & 1.26
$\pm$ 0.14\\ 
\hline
\hline
\end{tabular}
\label{t:fit}
\end{table}

\clearpage

\begin{thebibliography}{9}

\bibitem{dglap} Yu. L. Dokshiter, Sov. Phys. JETP 46 (1977) 641;\\
V. N. Gribov and L. N. Lipatov, Sov. J. Nucl. Phys. 15 (1972) 438 and
675;\\
G. Altarelli and G. Parisi, Nucl. Phys. 126 (1977) 297.
%
\bibitem{bfkl} E. A. Kuraev, L. N. Lipatov and V. S. Fadin,
  Sov. Phys. JETP 45 (1972) 199;\\
Y. Y. Balitsky and L. N. Lipatov, Sov. J. Nucl. Phys. 28 (1978) 822.
%
\bibitem{lipatov} V. S. Fadin and L. N. Lipatov, Phys. Lett. B429
  (1998) 127.
%
\bibitem{h1} H1 Collab., C. Adolff et al. DESY preprint DESY 98-143,
 hep-ex/9809028 (1998).
%
\bibitem{zeus} ZEUS Collab., J. Breitweg et al., DESY preprint DESY
  98-050, hep-ex/9805016 (1998).
%
\bibitem{mueller} A. H. Mueller, Nucl. Phys. B (Proc. Suppl.) 18 C
  (1991) 125.
%
\bibitem{bdm} J. Bartels, A. De Roeck and M. Loewe, Z. Phys. C 54
  (1992) 635.
%
\bibitem{tang} W--K. Tang, Phys. Lett. B 278 (1992) 363.
%
\bibitem{kms} J. Kwiecinski, A. D. Martin and P. J. Sutton,
  Phys. Rev. D 46 (1992) 921.
%
\bibitem{ariadne} L. L\"onnblad, Comp. Phys. Comm. 71 (1992) 15.
%
\bibitem{grvlo} M. G\"uck, E. Reya and A. Vogt, Z. Phys. C53 (1992) 127.
%
\bibitem{grvho} M. G\"uck, E. Reya and A. Vogt, Z. Phys. C67 (1995) 433.
%
\bibitem{cteq} H.L. Lai et al., Phys.Rev. D 55 (1997) 1280.
%
\bibitem{mrsr1} A. D. Martin, R. G. Roberts and W. J. Stirling,
  Phys. Lett. B 387 (1996) 419. 
%
\bibitem{royon} H. Navelet, R. Peschanski, Ch. Royon and S. Wallon,
  Phys. Lett. b 385 (1996) 357.
%
\bibitem{lotter} J. Bartels and H. Lotter, Phys. Lett. B 309 (1993) 400.
%
\end{thebibliography}
\end{document}